\documentclass[american,english,aps,preprint]{revtex4}
\usepackage[T1]{fontenc}
\usepackage[latin9]{inputenc}
\usepackage{amsmath}
\usepackage{amssymb}
\usepackage{graphicx}

\makeatletter
\@ifundefined{textcolor}{}
{%
 \definecolor{BLACK}{gray}{0}
 \definecolor{WHITE}{gray}{1}
 \definecolor{RED}{rgb}{1,0,0}
 \definecolor{GREEN}{rgb}{0,1,0}
 \definecolor{BLUE}{rgb}{0,0,1}
 \definecolor{CYAN}{cmyk}{1,0,0,0}
 \definecolor{MAGENTA}{cmyk}{0,1,0,0}
 \definecolor{YELLOW}{cmyk}{0,0,1,0}
}

\usepackage{babel}

\makeatother

\usepackage{babel}
\begin{document}

\title{Higher-Derivative Wess-Zumino Model in Three Dimensions}

\author{E. A. Gallegos}

\email{egallegoscollado@gmail.com}

\affiliation{Departamento de Física, Universidade Federal de Santa Catarina, Campus
Trindade, 88040-900, Florianópolis, SC, Brazil}

\author{C. R. Senise Jr.}

\email{carlossenise@unipampa.edu.br}

\affiliation{Universidade Federal do Pampa, Avenida Pedro Anunciação S/N, Vila
Batista, 96570-000, Caçapava do Sul, RS, Brazil}

\author{A. J. da Silva}

\email{ajsilva@fma.if.usp.br }

\affiliation{Instituto de Física, Universidade de São Paulo,  Caixa Postal 66318,
05315-970, São Paulo, SP, Brazil }
\begin{abstract}
We deform the well-known three dimensional $\mathcal{N}=1$ Wess-Zumino
model by adding higher derivative operators (Lee-Wick operators) to
its action. The effects of these operators are investigated both at
the classical and quantum levels.
\end{abstract}
\maketitle

\section{Introduction}

Higher derivative operators produce negative and positive effects
in quantum field theories. Among the negative effects are the lack
of unitary of the $S-$matrix, the presence of negative-norm states
(ghosts) and the violation of the Lorentz invariance. The ghosts,
strictly speaking, become evident by reformulating a given higher
derivative theory in terms of standard (lower derivative) operators,
i.e., by removing from it the higher derivative operators by means
of auxiliary fields. The lower-derivative theory obtained in this
manner is a theory with indefinite metric where the ``auxiliary''
fields play the role of ghost fields. In addition, this lower-derivative
reformulation of a higher derivative theory is a key step in its canonical
quantization based on Ostrogradski's approach \cite{Ostrogradski (1850)}.
A positive effect, on the other hand, is the improvement of the ultraviolet (UV)
behavior due to the exchange of both negative- and positive-norm states
in the Feynman integrals. Another way to say this is that the propagators
in theories with higher derivative kinetic operators are
more convergent at the UV limit $k^{2}\rightarrow\infty$
than in the usual ones, implying a better UV behavior of the Feynman
integrals (see \cite{Nakanishi (1971-2)}, for a review on these issues).
However, as reported in \cite{Antoniadis-Dudas-Ghilencea (2007)},
such positive effect does not always occur due to subtle problems
with the analytical continuation from the Minkowski to the Euclidean
space.

To circumvent or eliminate the negative effects and take advantage
of the positive ones, it is important to recognize that they have
its origins in the additional degrees of freedom introduced by the
higher derivative operators. Therefore, to construct a quantum field
theory with higher derivatives or with indefinite metric which satisfies
the minimal physical requirements (unitarity, Lorentz invariance and
positive-energy spectrum), it is necessary to devise suitable mechanisms
to get rid of ``runaway'' solutions or troublesome degrees of freedom.
Evidently, one means is to impose constraints or boundary conditions
on certain sectors of the theory.

A long time ago, Lee and Wick showed in \cite{Lee & Wick (1969),Lee & Wick (1970)}
that it is possible to construct a quantum field theory with indefinite
metric (or in another language, with higher derivative terms) in which
the $S-$matrix is relativistic and unitary. Specifically, they proposed
a variant of quantum electrodynamics (QED), which is the result of
introducing ``heavy'' negative-norm fields (heavy ghosts) in the
gauge and fermion sectors of the original QED theory, free of divergences.
The unitary problem, as a result of the presence of ghosts in Lee-Wick
QED, is avoided by requiring (as a boundary condition) that ghosts
do not belong to the asymptotic states of the $S-$matrix, a condition
that is possible only if the ghosts are unstable particles (i.e.,
if they have a non-vanishing decay width). On the other hand, as pointed
out by Lee and Wick \cite{Lee & Wick (1971)} afterward, the violation
of Lorentz invariance overlooked in Lee-Wick QED (see e.g. \cite{Nakanishi (1971-2)})
can also be avoided by adopting the prescription of Cutkosky et al.
\cite{Cutkosky (1969)} in the choice of the Feynman contours. Nonetheless,
in spite of these achievements, the Lee-Wick QED theory is still plagued
by some residual ghost effects such as acausality \cite{Coleman (1970)}.

Nowadays, higher derivative operators are common in several branches
of quantum physics and their negative effects are adequatedly dealt
with by applying the Lee-Wick's ideas or any other killing-ghost mechanism,
for instance, the method of ``perturbative constraints'' in nonlocal
theories (see, e.g., the work by Simon in \cite{Simon (1990)}). In
lower-energy effective theories, for example, higher derivative terms
occur naturally as a result of integrating out the massive fields
of a more fundamental theory and truncating its perturbative expansion,
while in gravity \cite{gravity-theory} such terms (quadratic or higher
powers of the curvature tensor) are generated dynamically by radiative
corrections. Higher derivative operators were also studied in susy/sugra
models, string theory, Randall-Sundrum models, cosmology, phase transitions
and Higgs models, and in other contexts (see \cite{Antoniadis-Dudas-Ghilencea (2007),Antoniadis & co-workes (2008)}
and references therein). More recently, the Lee-Wick's ideas were
applied in the framework of the standard model (SM) in order to solve
the hierarchy problem. Indeed, it was shown in \cite{Grinstein & co-workes (2008)}
that all quadratic divergent radiative corrections to the Higgs mass
are completely removed by introducing higher derivative kinetic terms
in each sector of the SM.

In this paper we bring together the good features (in relation to
the improved UV behavior) of supersymmetry and higher derivative operators.
In particular, we show that a single higher derivative kinetic operator
inserted in the usual Wess-Zumino action is sufficient to remove,
under a rather general assumption on the complex poles, all the susy
remaining divergences of the two-loop scalar self-energy. Notice that
in this kind of theory (where supersymmetry and higher derivative
properties are combined) two completely different mechanisms of removing
UV divergences are involved. The cancellation of UV divergences in
higher derivative theories occurs due to the exchange of normal and
ghost states, while in susy theories this cancellation is achieved
by the exchange of virtual particles with opposite statistics. Here
we show explicitly in connection with the two-loop self-energy how
both mechanisms work together to give a finite result.

On the other hand, to obtain some insights into the vacuum structure
of our higher derivative Wess-Zumino model, we compute the effective
potential at one-loop in the superfield formalism. By implicitly imposing
the susy condition of non-negative energy to throw away ``runaway''
solutions, we find that supersymmetry remains intact at one-loop order,
while the rotational symmetry is (spontaneously) broken iff a specific
condition for $\sigma_{1}\neq0$ is satisfied.

Our paper is organized as follows. In Sec. \ref{sec:HWZ3} we discuss
in general terms the three dimensional Wess-Zumino model and define
our higher derivative Wess-Zumino model (HWZ$_{3}$). This model results
of introducing three types of higher derivative operators in the usual
Wess-Zumino action. In Sec. \ref{sec:Effec-Pot} we compute the effective
potential at one-loop and study its minima. As mentioned above, it
is shown that supersymmetry remains intact at this order, while the
rotational symmetry defines two phases of the theory. In Sec. \ref{sec:Selft-Energy}
we analyze the UV behavior of the scalar self-energy up to
two-loops. Here we explicitly show how the mutual cancellation of
UV divergences happens. Finally our conclusions are given in Sec \ref{sec:Concl}.


\section{Higher-derivative Wess-Zumino Model\label{sec:HWZ3}}

Our starting point is the three-dimensional $\mathcal{N}=1$ Wess-Zumino
model,
\begin{equation}
S=\int d^{5}z\left\{ -\frac{1}{2}D^{\alpha}\overline{\Phi}D_{\alpha}\Phi+m\,\overline{\Phi}\Phi-U\right\} ,\label{eq:2.1}
\end{equation}
where $\Phi$ denotes a complex scalar superfield whose $\theta$-Taylor
expansion is
\begin{equation}
\Phi(x,\theta)=\varphi(x)+\theta^{\alpha}\psi_{\alpha}(x)-\theta^{2}F(x).\label{eq:2.2}
\end{equation}
Here $\varphi(x)$ and $F(x)$ represent bosonic fields and $\psi_{\alpha}(x)$
represents a fermionic field. Throughout the paper, we shall adopt
the notation of \cite{Gates-etal}. 

The superpotential $U$ in general involves terms of the form $\left(D_{\alpha}\right)^{N_{D}}\left(\overline{\Phi}\Phi\right)^{\frac{N_{\Phi}}{2}}$,
where $D_{\alpha}\doteq\partial_{\alpha}+i\theta^{\beta}\partial_{\alpha\beta}$
denotes the susy covariant derivative, $N_{D}$ and $N_{\Phi}$ are
respectively the number of susy derivatives ($D_{\alpha}$) and the
number of complex scalar superfields in a typical interaction vertex.

Note that the rotational symmetry $\Phi'(z)=e^{i\alpha}\Phi\left(z\right)$,
where $\alpha$ is a constant phase, and the Lorentz symmetry of the
Wess-Zumino model (\ref{eq:2.1}) restrict the values of $N_{D}$
and $N_{\Phi}$. In fact, Lorentz invariance requires an even number
of susy covariant derivatives, with their spinor indices completely
contracted, while rotational invariance requires an even number of
$\Phi$ and $\overline{\Phi}$ superfields. Consequently, $N_{D}$
and $N_{\Phi}$ must be even numbers. In addition, $N_{\Phi}$ has
to be greater than zero: $N_{\Phi}\geq2$.

By imposing the power-counting renormalizability condition, it is
possible to find the general form of $U$. To this end, we compute
the superficial degree of divergence $\omega$ of a typical Feynman
diagram. From the free superpropagator 
\begin{equation}
\left\langle \Phi\left(k,\,\theta\right)\overline{\Phi}\left(-k,\,\theta'\right)\right\rangle =-i\,\frac{D^{2}-m^{2}}{k^{2}+m^{2}}\delta^{2}\left(\theta-\theta'\right),\label{eq:2.3}
\end{equation}
which is obtained by inverting the kernel of the Wess-Zumino action
(\ref{eq:2.1}), and taking into account the Grassmann reduction procedure
\cite{Gates-etal}, one may show that
\begin{equation}
\omega=I\, V+2-\frac{E}{2}-\frac{n_{D}}{2},\label{eq:2.4}
\end{equation}
where $I$ denotes the index of divergence,
\begin{equation}
I\doteq\frac{N_{D}+N_{\Phi}}{2}-2,\label{eq:2.5}
\end{equation}
$V$ is the number of vertices in a typical Feynman diagram, $E$
is the number of external lines and $n_{D}$ is the number of susy
derivatives transferred to the external lines 

Since the index $I$ depends only on the form of the interaction vertex,
a simple analysis of (\ref{eq:2.4}) shows that the Wess-Zumino theory
is renormalizable when the condition $I\leq0$ is satisfied. This
in turn means that $U$ can be of three types:
\begin{equation}
U\propto(\overline{\Phi}\Phi)^{2},\qquad U\propto D^{\alpha}\overline{\Phi}D_{\alpha}\Phi,\qquad U\propto\overline{\Phi}\Phi,\label{eq:Types-Oper}
\end{equation}
thereby the only genuine vertex is the first one, while the others
(kinetic- and mass-like vertices) can be re-summed in the quadratic
part of the Wess-Zumino action (\ref{eq:2.1}) and then completely
absorbed by suitably redefining the superfield $\Phi$ and the parameters
of the theory.

Our goal now is to extend the Wess-Zumino model (\ref{eq:2.1}) by
adding to it higher derivative operators both at the kinetic and interaction
parts. These operators must fulfill all the symmetries of the conventional
theory, that is to say, they must be susy, Lorentz, and rotational
invariants. Furthermore, all must be Hermitian. There is indeed an
infinity of such operators, but in this work we shall restrict ourselves
to the following ones:
\begin{equation}
D^{2}\overline{\Phi}D^{2}\Phi,\qquad D^{\alpha}\overline{\Phi}\square D_{\alpha}\Phi,\qquad\left[\overline{\Phi}^{2}\left(D^{2}\Phi\right)^{2}+\Phi^{2}\left(D^{2}\overline{\Phi}\right)^{2}\right],
\end{equation}
where the two formers modify the kinetic part of the Wess-Zumino model,
while the latter modifies its interaction part.

Including these higher derivative operators in (\ref{eq:2.1}), with
$U=g\left(\overline{\Phi}\Phi\right)^{2}$, our Higher-derivative
Wess-Zumino model in three dimensions (HWZ$_{3}$) is described by
\begin{eqnarray}
S & = & \int d^{5}z\biggl\{-\frac{1}{2}D^{\alpha}\overline{\Phi}D_{\alpha}\Phi+a\, D^{2}\overline{\Phi}D^{2}\Phi-\frac{b}{2}D^{\alpha}\overline{\Phi}\square D_{\alpha}\Phi+m\,\overline{\Phi}\Phi-g\left(\overline{\Phi}\Phi\right)^{2}\nonumber \\
 &  & +\lambda\left[\overline{\Phi}^{2}\left(D^{2}\Phi\right)^{2}+\Phi^{2}\left(D^{2}\overline{\Phi}\right)^{2}\right]\biggr\}.\label{eq:2.6}
\end{eqnarray}
It is very simple to check out that the mass dimensions of the coefficients
in this action are $\left[a\right]=-1$, $\left[b\right]=-2$, $\left[m\right]=1$,
$\left[g\right]=0$ and $\left[\lambda\right]=-2$. Thus we shall
take $a=1/M$ and $b=1/\widetilde{M}^{2}$, where $M$ and $\widetilde{M}$
are in principle arbitrary and different mass parameters. However,
as we are regarding the higher derivative operators as the residual
(low-energy limit) effects of an underlying fundamental theory, it
is important to keep in mind that $a$, $b$ and $\lambda$ must be
very small compared with the original parameters of the theory. In
particular, $M$ and $\widetilde{M}$ must be of the Planck mass order.
It should also be noted that the $\lambda$ interaction is non-renormalizable
within the usual Wess-Zumino theory, according to the previous discussion.

In terms of the component fields, the HWZ$_{3}$ Lagrangian is given
by
\begin{equation}
\mathcal{L}=\mathcal{L}_{bos}+\mathcal{L}_{fer}+\mathcal{L}_{int},\label{eq:2.7}
\end{equation}
where
\begin{equation}
\mathcal{L}_{bos}=\overline{\varphi}\square\varphi+m\left(\varphi\overline{F}+\overline{\varphi}F\right)+F\overline{F}+a\left(\varphi\square\overline{F}+\overline{\varphi}\square F\right)+b\left(\overline{\varphi}\square^{2}\varphi+\overline{F}\square F\right),\label{eq:2.7a}
\end{equation}
\begin{equation}
\mathcal{L}_{fer}=\overline{\psi}i\partial\psi+m\overline{\psi}\psi+a\overline{\psi}\square\psi+b\overline{\psi}\square i\partial\psi,\label{eq:2.7b}
\end{equation}
and
\begin{eqnarray}
\mathcal{L}_{int} & = & -2g\left[\varphi^{2}\overline{\psi}^{2}+\overline{\varphi}^{2}\psi^{2}+\overline{\varphi}\varphi\left(\varphi\overline{F}+\overline{\varphi}F+2\overline{\psi}\psi\right)\right]+2\lambda\Bigl[\overline{\varphi}F\left(\overline{F}F+2\overline{\psi}i\partial\psi\right)\nonumber \\
 &  & +\psi^{2}\overline{F}^{2}+\overline{\varphi}^{2}\left(F\square\varphi-\left(\partial\psi\right)^{2}\right)+\mbox{h.c.}\Bigr].\label{eq:2.7c}
\end{eqnarray}
Here the contraction of spinor indices follows the north-west rule
($\searrow$) and the square of a spinor includes a factor of $1/2$
in its definition. So $\psi^{2}\doteq\frac{1}{2}\psi^{\alpha}\psi_{\alpha}$
and $\overline{\psi}i\partial\psi\doteq\overline{\psi}^{\alpha}i\partial_{\alpha}^{\;\beta}\psi_{\beta}$,
for instance.

By setting up the $F,\,\overline{F}$ equations of motion of the free
(i.e. switch off the $g$ and $\lambda$ couplings) part of (\ref{eq:2.7}),
\begin{eqnarray}
\left(a\square+m\right)\varphi+\left(b\square+1\right)F & = & 0\nonumber \\
\left(a\square+m\right)\overline{\varphi}+\left(b\square+1\right)\overline{F} & = & 0,\label{eq:2.8}
\end{eqnarray}
one can easily see that $F$ and $\overline{F}$ become dynamical fields
only when $b$ is different from zero. In fact, by taking $b=0$ in (\ref{eq:2.7}), $F$ and $\overline{F}$
play the role of auxiliary Lee-Wick fields by introducing the higher
derivative operator for the scalar $\varphi$ field only in the so-called
on-shell case.


\section{The effective potential at one-loop \label{sec:Effec-Pot}}

In what follows we compute the one-loop effective potential in order
to study the vacuum effect of the higher derivative operators. As
is well-known, the classical potential is given by the negative of
the classical HWZ$_{3}$ action (\ref{eq:2.7}) evaluated at constant
fields. Hence, writing the fields $\varphi$ and $F$ as
\begin{equation}
\varphi\doteq\frac{1}{\sqrt{2}}\left[\sigma_{1}+i\,\pi_{1}\right]\qquad\qquad F\doteq\frac{1}{\sqrt{2}}\left[\sigma_{2}+i\,\pi_{2}\right],\label{eq:2.9}
\end{equation}
where $\sigma_{i}$ and $\pi_{i}$ are real constant fields, the classical
potential $V_{cl}$ is given by
\begin{equation}
V_{cl}=-\frac{1}{2}\left(\sigma_{2}^{2}+\pi_{2}^{2}\right)-\left(\sigma_{1}\sigma_{2}+\pi_{1}\pi_{2}\right)\left[m-g\left(\sigma_{1}^{2}+\pi_{1}^{2}\right)+\lambda\left(\sigma_{2}^{2}+\pi_{2}^{2}\right)\right].\label{eq:2.10}
\end{equation}
Since the effective potential $V$ possesses the rotational SO(2)
symmetry, $\sigma_{i}'+i\pi_{i}'=\exp\left(i\,\omega\right)\left(\sigma_{i}+i\pi_{i}\right)$,
that inherits from the classical HWZ$_{3}$ action, we shall take
$\pi_{i}=0$ in (\ref{eq:2.10}) to simplify the analysis of the vacuum
structure of the theory. Doing this and solving the Euler-Lagrange
equation $\partial V_{cl}/\partial\sigma_{2}=0$, one finds two solutions
for $\sigma_{2}$: 
\begin{equation}
\sigma_{2}^{\pm}=\frac{-1\pm\sqrt{1+12\lambda\triangle}}{6\lambda\sigma_{1}},\label{eq:2.11}
\end{equation}
where $\triangle=g\,\sigma_{1}^{2}\left(\sigma_{1}^{2}-r\right)$,
with $r=m/g$. As we shall see below $r$ is an order-parameter related
to the spontaneous breakdown of the rotational symmetry at the classical
level.

\begin{figure}
\begin{centering}
\includegraphics[scale=0.45]{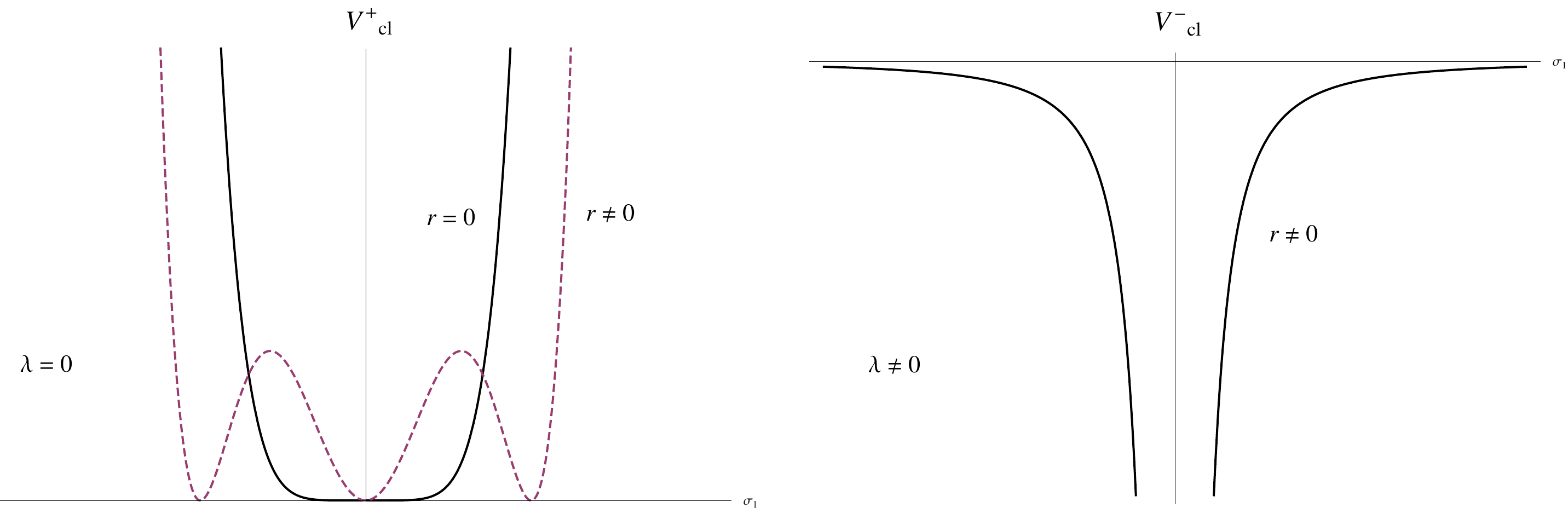}
\par\end{centering}

\caption{\label{fig:ClassPot}Typical curves of $V_{cl}^{+}$ and $V_{cl}^{-}$.
In $V_{cl}^{+}$ graph, with $\lambda=0$, the solid line represents
the unbroken phase ($r=0$), while the dashed line the broken phase
($r\neq0$) of the rotational symmetry. In $V_{cl}^{-}$ graph, the
solid line exhibits the behavior of an atypical potential curve.}
\end{figure}

Eliminating the ``auxiliary'' field $\sigma_{2}$ of the classical
potential $V_{cl}$ by means of (\ref{eq:2.11}), one obtains
\begin{equation}
V_{cl}^{\pm}=\frac{1}{108\lambda^{2}\sigma_{1}^{2}}\left[-1-18\lambda\triangle\pm\left(1+12\lambda\triangle\right)^{\frac{3}{2}}\right].
\end{equation}
In order to analyze the physical implications of this expression,
let us consider $V_{cl}^{+}$ and $V_{cl}^{-}$ separately. Taking
the positive potential $V_{cl}^{+}$ and expanding it around $\lambda=0$,
\begin{equation}
V_{cl}^{+}=\frac{g^{2}\sigma_{1}^{2}}{2}\left(\sigma_{1}^{2}-r\right)^{2}\left[1-2\lambda\triangle+9\lambda^{2}\triangle^{2}+\mathcal{O}\left(\lambda^{3}\right)\right],\label{eq:2.12a}
\end{equation}
we find out that $V_{cl}^{+}$ is analytic at $\lambda=0$ and positive
definite $\left(V_{cl}^{+}\geq0\right)$, as required by supersymmetric
grounds. Fig. \ref{fig:ClassPot} displays two characteristic curves
of $V_{cl}^{+}$ for $r=0$ and $r\neq0$, setting $\lambda=0$. They
show us that there are two phases associated to the rotational symmetry:
an unbroken phase of the rotational symmetry corresponding to $r=0$
and an spontaneous broken phase corresponding to $r\neq0$. Classically,
note also that supersymmetry remains intact at both phases. An analysis
for the $\lambda\neq0$ case leads us to the same conclusions.

Now taking the negative potential $V_{cl}^{-}$ and its expansion
around $\lambda=0$,
\begin{equation}
V_{cl}^{-}=-\frac{1}{54\lambda^{2}\sigma_{1}^{2}}-\frac{g\left(\sigma_{1}^{2}-r\right)}{3\lambda}-\frac{g^{2}\sigma_{1}^{2}\left(\sigma_{1}^{2}-r\right)}{2}\left[1-2\lambda\triangle+9\lambda^{2}\triangle^{2}+\mathcal{O}\left(\lambda^{3}\right)\right],\label{eq:2.12b}
\end{equation}
we see that in this case $V_{cl}^{-}$ is singular at $\lambda=0$
and negative definite without a lower-energy bound (see Fig. 1). So,
from a physical point of view, $V_{cl}^{+}$ is the only acceptable
potential that is positive definite and bounded from below. 

Next we shall compute the one-loop contribution to the effective potential
by employing the steepest-descent method \cite{Itzykson-Zuber (1980)}
or, as it is also known, the Jackiw functional method \cite{Jackwi (1974)},
implemented in the superspace. For simplicity and technical reasons,
from now on we will set $b=0=\mbox{\ensuremath{\lambda}}$. The one-loop
contribution is given by
\begin{equation}
V_{1}=-\frac{i}{2v}\mbox{ln\,\ Det}\left(A-B\right)-\frac{i}{2v}\mbox{ln\,\ Det}C,\label{eq:2.13}
\end{equation}
where $v=\int d^{3}x$,
\begin{eqnarray}
A(z,\, z') & = & \left[m+a\square-g\left(3\sigma_{1}^{2}+\pi_{1}^{2}\right)+D^{2}+2g\left(3\sigma_{1}\sigma_{2}+\pi_{1}\pi_{2}\right)\theta^{2}\right]\delta^{5}\left(z-z'\right),\nonumber \\
B\left(z,\, z'\right) & = & \int dz_{1}dz_{2}H\left(z,\, z_{1}\right)C^{-1}\left(z_{1},\, z_{2}\right)H\left(z_{2},\, z'\right),\\
C\left(z,\, z'\right) & = & \left[m+a\square-g\left(\sigma_{1}^{2}+3\pi_{1}^{2}\right)+D^{2}+2g\left(\sigma_{1}\sigma_{2}+3\pi_{1}\pi_{2}\right)\theta^{2}\right]\delta^{5}\left(z-z'\right),\nonumber 
\end{eqnarray}
with 
\begin{equation}
H\left(z,\, z'\right)=\left[-2g\sigma_{1}\pi_{1}+2g\left(\sigma_{1}\pi_{1}+\sigma_{2}\pi_{2}\right)\theta^{2}\right]\delta^{5}\left(z-z'\right).
\end{equation}
The superspace functional determinants which appear in $V_{1}$ are
evaluated using the $\zeta$- functional method \cite{Burgess,Roberto-Adilson},
taking advantage of the rotational SO(2) symmetry (i.e. setting $\pi_{i}=0$).
The result can be cast in the form{\small 
\begin{equation}
V_{1}=\frac{1}{2}\int\frac{d^{3}k}{\left(2\pi\right)^{3}}\left\{ \mbox{ln}\left[\frac{a^{2}\left(k^{2}\right)^{2}+\alpha(3)k^{2}+\beta\left(3,3\right)}{a^{2}\left(k^{2}\right)^{2}+\alpha\left(3\right)k^{2}+\beta\left(3,0\right)}\right]+\mbox{ln}\left[\frac{a^{2}\left(k^{2}\right)^{2}+\alpha\left(1\right)k^{2}+\beta\left(1,1\right)}{a^{2}\left(k^{2}\right)^{2}+\alpha\left(1\right)k^{2}+\beta\left(1,0\right)}\right]\right\} ,
\end{equation}
}where, defining $\omega\left(n\right)\doteq m-n\cdot g\sigma_{1}^{2}$
,
\begin{equation}
\alpha\left(n\right)\doteq1-2a\,\omega\left(n\right)\qquad\mbox{and}\qquad\beta\left(n_{1},n_{2}\right)\doteq\omega\left(n_{1}\right)^{2}+n_{2}\cdot\left(2g\sigma_{1}\sigma_{2}\right).
\end{equation}

Note that setting $a=0$ this result reduces to the usual case where
there are no ghosts in the theory, since the $\left(k^{2}\right)^{2}$
terms vanish. 

In the following let us focus on the case with $a\neq0$. To solve
the above integrals we express each numerator and denominator as the
product of two binomial factors of the form $(k^{2}+M_{i}^{2})$ and
employ the formulas in \cite{Tan-etal}. In this way, we get
\begin{equation}
V_{1}=-\frac{1}{12\pi}\sum_{i=+,-}\left[M_{i}\left(3,\,3\right)^{3}-M_{i}\left(3,\,0\right)^{3}+M_{i}\left(1,\,1\right)^{3}-M_{i}\left(1,\,0\right)^{3}\right]\label{eq:2.14}
\end{equation}
where
\begin{equation}
M_{\pm}\left(n_{1},\, n_{2}\right)\doteq\frac{1}{\sqrt{2}\, a}\sqrt{1-2a\omega\left(n_{1}\right)\pm\sqrt{1-4a\omega\left(n_{1}\right)-n_{2}\cdot\left(8a^{2}g\sigma_{1}\sigma_{2}\right)}}.
\end{equation}
Adding (\ref{eq:2.10}), with $\pi_{i}=0=\lambda$, and (\ref{eq:2.14}),
we get the one-loop effective potential:{\small 
\begin{equation}
V_{eff}=-\frac{\sigma_{2}^{2}}{2}-\sigma_{1}\sigma_{2}\left(m-g\sigma_{1}^{2}\right)-\frac{1}{12\pi}\sum_{i=+,-}\left[M_{i}\left(3,\,3\right)^{3}-M_{i}\left(3,\,0\right)^{3}+M_{i}\left(1,\,1\right)^{3}-M_{i}\left(1,\,0\right)^{3}\right].
\end{equation}
}At first glance, this result seems fairly intricate, however, considering
the $\sigma_{2}-$linear approximation for the one-loop contribution,
we will get some light about its physical implications. In three dimensions
$\mathcal{N}=1$ susy, this approximation is enough to analyze the
possibility of susy breaking \cite{Burgess}. Thus, the one-loop effective
potential is given by 
\begin{equation}
V_{eff}=-\frac{\sigma_{2}^{2}}{2}-\sigma_{1}\sigma_{2}\left[m-g\sigma_{1}^{2}+X\left(a,\,\sigma_{1}\right)\right],
\end{equation}
{\small where}
\begin{equation}
X\left(a,\,\sigma_{1}\right)\doteq\frac{g}{4\pi}\left[\frac{M_{-}\left(1,\,0\right)-M_{+}\left(1,\,0\right)}{\sqrt{1-4\, a\,\omega\left(1\right)}}+\frac{3\left[M_{-}\left(3,\,0\right)-M_{+}\left(3,\,0\right)\right]}{\sqrt{1-4\, a\,\omega\left(3\right)}}\right].
\end{equation}
{\small After eliminating the auxilary field $\sigma_{2}$ by means
of its equation of motion, one gets}
\begin{equation}
V_{eff}=\frac{1}{2}\sigma_{1}^{2}\left[m-g\sigma_{1}^{2}+X\left(a,\,\sigma_{1}\right)\right]^{2}.
\end{equation}
{\small Notice that $V_{eff}\geq0$ and its minima occur at $\sigma_{1}=0$
or $\sigma_{1}\neq0$, provided that $\sigma_{1}$ is the solution
to the equation:}
\begin{equation}
g\sigma_{1}^{2}=m+X\left(a,\,\sigma_{1}\right).\label{eq:susy-condition}
\end{equation}
This result means that susy is unbroken at one-loop order, while the
rotational symmetry is spontaneously broken at this order only if
there is a non-trivial solution $\sigma_{1}\neq0$ for the condition
(\ref{eq:susy-condition}).

On the other hand, the series expansion for $V_{eff}$ around $a=0$,
\begin{equation}
V_{eff}\approx\frac{g^{2}\sigma_{1}^{2}}{2\pi^{2}a^{2}}+\frac{g\sigma_{1}^{2}\left|m-g\sigma_{1}^{2}\right|}{\pi a}+\frac{1}{2}\sigma_{1}^{2}\left(m-g\sigma_{1}^{2}\right)^{2}+\mathcal{O}\left(a^{10}\right),
\end{equation}
shows us that $V_{eff}$ is singular at $a=0$. This result is very
important and shows that a higher derivative term with a small coefficient,
no matter how small it is, cannot be treated as a perturbation in
the lower-derivative theory. Indeed, this result is well known in
non-susy theories with higher derivative operators studied long ago in the
literature \cite{Simon (1990)}.


\section{Ultraviolet analysis of the self-energy up to two-loops\label{sec:Selft-Energy}}

In this section, we analyze the UV behavior of the scalar
self-energy for the HWZ$_{3}$ model. We show that the introduction
of a higher derivative operator in the Lagrangian (and consequently
in the propagator) improves the structure of the divergences that
appear in the momentum-space integrals, in contrast with the usual
Wess-Zumino model. In fact, we shall see that all the integrals appearing
in these corrections are finite in the UV regime, and so
we have a much better UV behavior when comparing with the usual
case.

\begin{figure}
\begin{centering}
\includegraphics[scale=0.6]{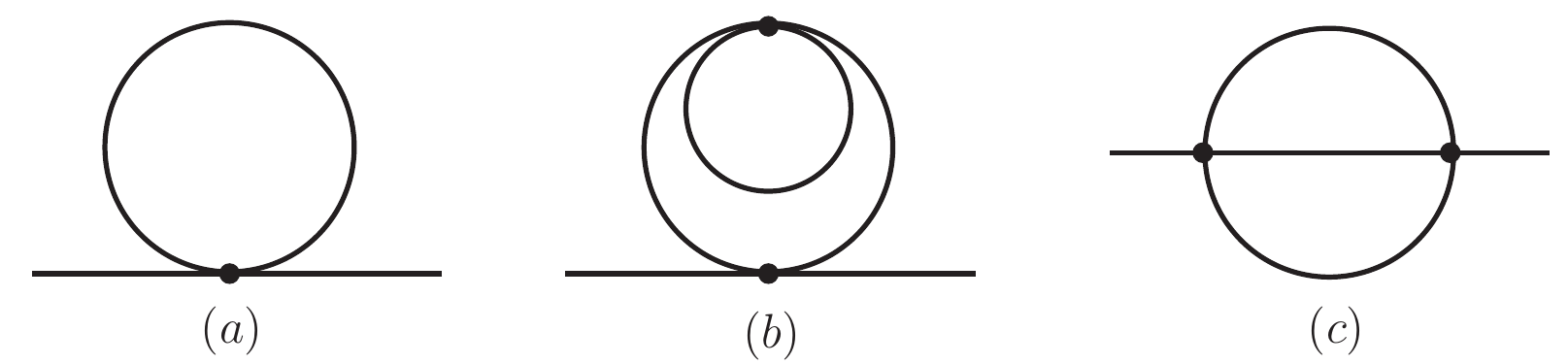}
\par\end{centering}

\caption{\label{fig:Self-energy}Self-energy radiative corrections up to two-loops
in the HWZ$_{3}$ model.}
\end{figure}

The Feynman diagrams which contribute to the scalar self-energy up to two-loops are depicted in Fig. \ref{fig:Self-energy}. As in the previous discussion about the effective potential at the one-loop order, here we set $b=0=\lambda$ as well. Hence the free propagator of the HWZ$_3$ model is given by
\begin{eqnarray}
\Delta(k,\theta-\theta^{\prime})&=&\langle0|T\Phi(k,\theta)\bar{\Phi}(-k,\theta^{\prime})|0\rangle  \nonumber\\
&=&\frac{i}{(m-ak^{2})^{2}+k^{2}}[m-ak^{2}-D^{2}]\delta^{2}(\theta-\theta^{\prime})  \nonumber\\
&=&i[A(k)+B(k)D^{2}]\delta^{2}(\theta-\theta^{\prime}) \ , 
\label{propb=0} 
\end{eqnarray}
with
\begin{equation}
A(k)=\frac{m-ak^{2}}{(m-ak^{2})^{2}+k^{2}} \ \ \ , \ \ \ B(k)=-\frac{1}{(m-ak^{2})^{2}+k^{2}} \ . 
\label{AB} 
\end{equation}

In three dimensions and within the dimensional reduction (DReD$_3$) scheme \cite{Siegel (1979-1980)}, the one-loop diagram of Fig. \ref{fig:Self-energy} is finite. This statement is very general and well-known in lower-derivative theories \cite{Avdeev-etal (1992-3)}  and is not spoiled by the introduction of higher derivative operators \cite{Grinstein & O'Connell (2008)}. Indeed, as far as the UV behavior of higher-derivative theories is concerned, the Wick rotation can be performed in the usual way disregarding all contributions of the complex poles in the complex energy plane. The reason is that in the majority of the theories with higher derivative operators the residue contributions of the complex poles in the Cauchy's residue theorem are finite and so the form of the energy contour in the Feynman integrals is irrelevant. Note however that in order to construct a relativistic and unitary $S-$matrix it is necessary to fix the Feynman contour acoording to the Cutkosky's prescription \cite{Cutkosky (1969)}.

Bearing these remarks in mind we proceed with the UV analysis of the diagram (b) in Fig. \ref{fig:Self-energy}. The expression for this diagram is written as 
\begin{equation}
{\mathcal G}_{2b} = {\mathcal C}_{2b} \!\int\!d\tilde{p}d\tilde{k}d\tilde{q} \!\int\!d^{2}\theta_{1}d^{2}\theta_{2} \Phi(\theta_1,p) \bar{\Phi}(\theta_1,-p)\Delta(k,\theta_1-\theta_2)\Delta(k,\theta_2-\theta_1)\Delta(q,\theta_2-\theta_2) \ , 
\label{G1} 
\end{equation}
where $p$ (with $d \tilde{p}=d^3p/(2\pi)^3$ and so on) is the external momentum, $k$ and $q$ are the momenta appearing in the loops and ${\mathcal C}_{2b}$ is a numerical constant and since we are concerned only with the divergent behavior of the diagram we do not worry about it. 
After carrying out the $D$-algebra, the expression above takes the form 
\begin{eqnarray} 
{\mathcal G}_{2b} & \propto & \!\int\!d\tilde{p}d\tilde{k}d\tilde{q}\!\int\!d^{2}\theta\left\{A(k)B(k)B(q)\bar{\Phi}(\theta,-p) \Phi(\theta,p)+[B(k)]^{2}B(q)\left[D^{2}\bar{\Phi}(\theta,-p)\Phi(\theta,p)\right.\right.\nonumber\\ &&\left.\left.\hspace{3cm}+D^{\alpha}\bar{\Phi}(\theta,-p)D_{\alpha}\Phi(\theta,p)+\bar{\Phi}(\theta,-p)D^{2}\Phi(\theta,p)\right]\right\} \ , 
\label{G1Dalgebra} 
\end{eqnarray} 
and as can be easily seen the integrals in the momenta $k$ and $q$ are independent, and so by the same argument given above each one of these one-loop integrals is finite in the DReD$_3$ scheme.

Now let us consider the last diagram in Fig. \ref{fig:Self-energy}. Its mathematical expression is given by
\begin{equation}
{\mathcal G}_{2c} = {\mathcal C}_{2c} \!\int\!d\tilde{p}d\tilde{k}d\tilde{q}\!\int\!d^{2}\theta_{1}d^{2} \theta_{2} \Phi(\theta_1,p) \bar{\Phi}(\theta_2,-p)\Delta(k,\theta_1-\theta_2)\Delta(r,\theta_1-\theta_2)\Delta(q,\theta_2-\theta_1) \ , 
\label{G2} 
\end{equation} 
where $r=p+k-q$ and ${\mathcal C}_{2c}$ is a numerical constant and we do not worry about it, in the same way that we did with the constant ${\mathcal C}_{2b}$ in diagram (b). 
After the $D$-algebra, one obtains 
\begin{eqnarray} 
{\mathcal G}_{2c} & \propto & \!\int\!d\tilde{p}d\tilde{k}d\tilde{q}\!\int\!d^{2}\theta\left\{[A(r)B(k)B(q)+A(q)B(k)B(r) +A(k)B(q)B(r)]\bar{\Phi}(\theta,-p)\Phi(\theta,p)\right.\nonumber\\ &&\left.\hspace{3cm}+B(k)B(q)B(r)D^{2}\bar{\Phi}(\theta,-p)\Phi(\theta,p)\right\} \ . 
\label{G2Dalgebra} 
\end{eqnarray}
The analysis of the UV behavior for this diagram is much more cumbersome. It presents the so-called ``overlapping momenta'', which did not appear in diagram (b). In order to isolate the divergent part from the finite part of the integral above, we must take its Taylor series expansion around $p^2=0$. Since each differentiation with respect to the external momentum improves the convergence of the integrand, the divergences will reside only in the first few terms of its Taylor expansion. In our case the Taylor expansion of (\ref{G2Dalgebra}) around $p^2=0$ is given by
\begin{equation}
{\mathcal G}_{2c} \propto \!\int\!d\tilde{p}\!\int\!d^{2}\theta[{\mathcal R}(p^2)D^{2}\bar{\Phi}(\theta,-p)\Phi(\theta,p)
+{\mathcal S}(p^2)\bar{\Phi}(\theta,-p)\Phi(\theta,p)] \ , 
\label{G2expanded} 
\end{equation}
where
\begin{equation}
{\mathcal R}(p^2)={\mathcal R}_{1}+p^{2}{\mathcal R}_{2}+ {\mathcal O}(p^4), 
\label{R}  
\end{equation} 
\begin{equation} 
{\mathcal S}(p^2)={\mathcal S}_{1}+p^{2}{\mathcal S}_{2}+{\mathcal O}(p^4)\ ,
\label{S}  
\end{equation}
with, defining $l\doteq k-q$ and $f(x)\doteq (m-a\,x)^2+x$,
\begin{eqnarray}
{\mathcal R}_{1}&=&\int\!d\tilde{k}d\tilde{q}\,\frac{r_1}{f(k^2)f(q^2)f(l^2)}, \,\nonumber\\
{\mathcal R}_{2}&=&\int\!d\tilde{k}d\tilde{q}\,\frac{r_{2a}l^2+r_{2b}f(l^2)+r_{2c}l^2f(l^2)}
{f(k^2)f(q^2)[f(l^2)]^3}, \,\nonumber\\
{\mathcal S}_{1}&=&\int\!d\tilde{k}d\tilde{q}\,\frac{s_{1a}+s_{1b}(k^{2}+q^{2}+l^{2})}{f(k^2)f(q^2)f(l^2)}, \,\nonumber\\
{\mathcal S}_{2}&=&\int\!d\tilde{k}d\tilde{q}\,\frac{[s_{2a}(l^2)^2+(s_{2b}+s_{2c}l^2)(k^2+q^2)]f(l^2)
+[P(l^2)+k\cdot q \, Q(l^2)]l^2}{f(k^2)f(q^2)[f(l^2)]^3}.\,\nonumber
\label{mom integrals RS}
\end{eqnarray}
Here $r_i$ and $s_i$ denote numerical constants which might depend on the parameters $a$ and $m$, while $P$ and $Q$ are polynomials in $l^2$ of degrees $4$ and $2$, respectively.

At this point two comments are in order. Firstly, since the two-loop self-energy in the usual Wess-Zumino model, i.e., without higher derivative operators, involves merely logarithmic and linear divergences (see Eq. (\ref{eq:2.4}), setting $I=0$), we focus our attention only to the first term in the Taylor expansions for ${\mathcal R}$ and ${\mathcal S}$. Clearly, these terms enclose all the divergences present in the two-loop self-energy of the HWZ$_3$ model. Secondly, we should point out that the $(k,q)-$ integrals associated with $p^{\mu}-$ linear terms in the Taylor expansions are identically zero, a result that is in agreement with the Lorentz invariance of the model.

As a result of the indefinite metric in Lee-Wick theories, in particular our HWZ$_3$ model, the ${\mathcal R}_i$ and ${\mathcal S}_i$ Feynman integrals turn out to be finite. This assertion can be proved explicitly by evaluating each integral and observing the mutual cancellation between the divergent contributions from positive- and negative-norm (ghost)  states, or implicitly by examining the superficial degree of divergence of each diagram.

As the algebraic manipulations required to solve the ${\mathcal R}_i$ and ${\mathcal S}_i$ integrals are very lengthy to be shown here, we are going to illustrate the divergence cancellation by evaluating in detail only the simplest integral, namely ${\mathcal R}_{1}$. Up to a constant, we first rewrite the integral as
\begin{eqnarray}
{\mathcal R}_{1}&=&\int\!d\tilde{k}d\tilde{q}\frac{1}{[(m-ak^{2})^{2}+k^{2}][(m-aq^{2})^{2}+q^{2}][(m-al^{2})^{2}+l^{2}]} \nonumber\\        &=&\int\!d\tilde{k}d\tilde{q}\frac{1}{a^{6}(k^{2}+\alpha)(k^{2}+\beta)(q^{2}+\alpha)(q^{2}+\beta)(l^{2}+\alpha)(l^{2}+\beta)} \ , 
\label{R1} 
\end{eqnarray}
with 
\begin{eqnarray*} 
\alpha,\beta=\frac{1}{2a^{2}}\left(1-2am\mp\sqrt{1-4am}\right) \ . 
\end{eqnarray*}
Next we use the method of partial fraction decomposition to write  
\begin{eqnarray} \frac{1}{(k^{2}+\alpha)(k^{2}+\beta)}=\frac{1}{\beta-\alpha}\left(\frac{1}{k^{2}+\alpha}-\frac{1}{k^{2}+\beta}\right) \ , \nonumber 
\end{eqnarray} 
and split the integrand into eight fractions, obtaining 
\begin{eqnarray}
{\mathcal  R}_{1}\!\! &=& \!\!\frac{1}{a^{6}(\beta\!-\!\alpha)^{3}} \int\!d\tilde{k}d\tilde{q}
\left[
\frac{1}{(k^{2}\!+\!\alpha)(q^{2}\!+\!\alpha)(l^{2}\!+\!\alpha)}\!-\!\frac{1}{(k^{2}\!+\!\alpha)(q^{2}\!+\!\alpha)(l^{2}\!+\!\beta)}\right. \nonumber\\ &&\left.-\frac{1}{(k^{2}\!+\!\alpha)(q^{2}\!+\!\beta)(l^{2}\!+\!\alpha)}\!+\!\frac{1}{(k^{2}\!+\!\alpha)(q^{2}\!+\!\beta)(l^{2}\!+\!\beta)}\!-\!\frac{1}{(k^{2}\!+\!\beta)(q^{2}\!+\!\alpha)(l^{2}\!+\!\alpha)}\right. \nonumber\\ &&\left.+\frac{1}{(k^{2}\!+\!\beta)(q^{2}\!+\!\alpha)(l^{2}\!+\!\beta)}\!+\!\frac{1}{(k^{2}\!+\!\beta)(q^{2}\!+\!\beta)(l^{2}\!+\!\alpha)}\!-\!\frac{1}{(k^{2}\!+\!\beta)(q^{2}\!+\!\beta)(l^{2}\!+\!\beta)}\right] \ . 
\label{R1partialfraction} 
\end{eqnarray}
Whitin the dimensional regularization scheme, each one of these integrals is calculated using the following formula \cite{Tan-etal}:
\begin{eqnarray} 
{\mathcal I}(x,y,z)&=&\int\frac{d^{D}k d^{D}q}{(2\pi)^{2D}}\frac{1}{(k^{2}+x)(q^{2}+y)(l^{2}+z)}\nonumber\\
&=&\frac{\mu^{-2\epsilon}}{32\pi^{2}}\left\{\frac{1}{\epsilon}-\gamma_{E}+1-\ln\left[\frac{(\sqrt{x}+\sqrt{y}+\sqrt{z})^{2}}{4\pi\mu^{2}}\right]\right\}+{\mathcal B_{res}} \ , 
\label{Ixyz} 
\end{eqnarray} 
where $\epsilon=3-D$, with $D$ labeling the dimension of the spacetime, $\gamma_{E}=0.5772$ is Euler's constant and ${\mathcal B_{res}}$ denotes the sum of residues of the integrand over all complex poles inside an energy contour appropriate for performing the ``Wick rotation'' (i.e. an energy contour which permits to change the real integration axis to the imaginary axis by means of the Cauchy's residue theorem). Strictly speaking, the analytical continuation from the Minkowski to the Euclidean space, which is common in conventional quantum field theories, is lost in the theories with  higher derivative operators (i.e. Lee-Wick theories) due to the presence of complex poles in the complex energy plane. This fact is reflected, for example, in the non-vanishing value of ${\mathcal B_{res}}$ that one finds in this sort of theory.

Using the formula (\ref{Ixyz}), we express ${\mathcal R}_1$ in the form
\begin{equation} 
{\mathcal R}_1 = \frac{1}{a^{6}(\beta-\alpha)^{3}}\left[{\mathcal I}(\alpha,\alpha,\alpha)
-3{\mathcal I}(\alpha,\alpha,\beta)+3{\mathcal I}(\alpha,\beta,\beta)-{\mathcal I}(\beta,\beta,\beta)\right] \ . 
\label{R1final} 
\end{equation}
From (\ref{Ixyz}) and (\ref{R1final}), and assuming without proof that ${\mathcal B_{res}}$ is finite, one can see that the divergent parts of the ${\mathcal I}$ terms cancel mutually, so that the integral ${\mathcal R}_1$ as a whole is in fact finite, as was claimed before. The same procedure can be used to calculate the other integrals and all of them happen to be finite. The finiteness of the self-energy in the HWZ$_3$ model is in contrast with the usual Wess-Zumino model, in which the diagram (c) in Fig. \ref{fig:Self-energy} gives a non-vanishing divergent contribution in the UV regime, showing us that a higher derivative kinetic operator improves the behavior of the model in this regime.

There is a more elegant and general form to see why this better UV behavior is achieved. This is the study of the superficial degree of divergence $\omega$ of a diagram. Recalling Eq.~(\ref{propb=0}), we can see that the two terms in the expression for the superpropagator, one proportional to $A(k)$ and the other proportional to $B(k)$, give different contributions to a given diagram, since $A(k)$ has power $-2$ and $B(k)$ has power $-4$ in the momentum $k$. Moreover, $B(k)$ is accompanied by a superspace derivative $D^{2}$, which is not the case for $A(k)$. We have therefore to consider that the complete propagator comprises two types of propagators in order to compute the superficial degree of divergence. These two propagators are defined by the $A(k)-$ and $B(k)-$terms in (\ref{propb=0}):
\begin{equation}
\Delta(k,\theta-\theta^{\prime})= \Delta_A(k,\theta-\theta^{\prime})+\Delta_B(k,\theta-\theta^{\prime}),
\end{equation}
where
\begin{eqnarray}
\Delta_A(k,\theta-\theta^{\prime})&=&\frac{i(m-ak^{2})}{(m-ak^{2})^{2}+k^{2}}\delta^2(\theta-\theta^{\prime})\nonumber\\
\Delta_B(k,\theta-\theta^{\prime})&=&-\frac{iD^2}{(m-ak^{2})^{2}+k^{2}}\delta^2(\theta-\theta^{\prime}).\nonumber
\end{eqnarray}
Taking into account the two types of propagators, it is straightforward to show that
\begin{equation} 
\omega=2-x, 
\label{omega} 
\end{equation}  
where
\begin{equation}
x= 2V+P_B+\frac{n_D}{2},
\end{equation}
and, as before, $V$ and $n_D$ are respectively the number of vertices and the number of susy derivatives transferred to the external lines, and $P_B$ is the number of propagators of the type $B$ in a given diagram. Since $x$ is strictly an integral positive number ($x \geq 1$), our HWZ$_3$ model exhibits only logarithmic ($\omega=0$) and linear ($\omega=1$) divergences.

On the other hand, from Eqs.~(\ref{G1Dalgebra}) and~(\ref{G2Dalgebra}) we see that $n_D$ assumes only two values, $n_D=0$ and $n_D=2$ and, for the two-loop diagrams in Fig.~\ref{fig:Self-energy}, we have that $V=2$. Thus, by considering the worst case (i.e. $P_B=0$) in both diagrams, one sees that $\omega<0$, showing that these diagrams are finite.


\section{Summary and Conclusions\label{sec:Concl}}

In this work we investigated the classical (and quantum) effects of
three types of higher derivative operators introduced in the Lagrangian
of the three-dimensional Wess-Zumino model. These operators respect
all the symmetries of the original model, but the potential $\lambda-$operator
turns out to be non-renormalizable by power-counting arguments. At
the classical level, we show that these Lee-Wick operators modify
the structure of the equations of motion for the component fields.
In particular, one finds that the Lee-Wick $b-$ operator promotes
the component field $F\left(\bar{F}\right)$ from an auxiliary field
to a dynamical field. 

We also considered the quantum aspects of the model in two distinct
analysis. First, we computed the one-loop correction to the classical
potential, which is singular at a null value for the higher derivative
parameter $a$. This fact was already expected \cite{Simon (1990)}
and shows that the higher derivative term cannot be treated as a perturbation
in the lower derivative theory. After this, we analyze the UV
behavior of the self-energy up to two-loops. We showed by direct computation
of one of the momentum-space integrals and also by calculating the
superficial degree of divergence that the ``setting-sun'' diagram
gives a finite contribution to the self-energy, in contrast with the
usual Wess-Zumino model. This explicitly shows that the introduction of the higher derivative operator improves the
behavior of the theory in the UV regime. 

As future efforts, we shall intend to analyze the features of the
model presented here with $b,\,\lambda\neq0$. This is a more involved
work, due to the fact that the integrals appearing in the one-loop
correction to the effective potential and in the correction to the
self-energy are much more cumbersome and, besides that, the number
of two-loop diagrams increases considerably. Also, we will consider
the study of three-dimensional supersymmetric models with gauge fields
within the framework of higher derivative models, a very interesting
task which deserves attention. The situation in this case is different
from what we have in the present work, for the gauge potential multiplet
is a spinorial function on superspace and the procedure to obtain
the propagators, even without higher derivative operators, is more
intricated \cite{Gallegos-Adilson}.


\begin{acknowledgments}
We would like to thank  R. V. Maluf for useful and fruitful discussions. This work was partially supported by Conselho Nacional de Desenvolvimento
Científico e Tecnológico (CNPq). The work by E. A. Gallegos has been
supported by CAPES-Brazil.
\end{acknowledgments}


\end{document}